\documentclass[11pt,dvips]{article}
\usepackage{epsfig,times}
\usepackage{picinpar}
\usepackage{wrapfig}
\usepackage{floatflt}
\makeatletter
\renewcommand{\section}{\@startsection{section}{1}{0in}
	{0.4\baselineskip}{0.1\baselineskip}{\Large\bf}}
\renewcommand{\subsection}{\@startsection{subsection}{2}{0in}
	{0.25\baselineskip}{-\baselineskip}{\large\bf}}
\renewcommand{\subsubsection}{\@startsection{subsubsection}{3}{0in}
	{0.1\baselineskip}{-\baselineskip}{\normalsize\bf}}
\makeatother
\def\beq{\begin{equation}}
\def\eeq{\end{equation}}

\def\ol{\overline}
\def\ul{\underline}
\begin{document}

%
\makeatletter\newcommand{\ps@icrc}{
\renewcommand{\@oddhead}{\slshape{HE.6.1.03}\hfil}}
\makeatother\thispagestyle{icrc}
%
\markright{HE.6.1.03}
%

%
\begin{center}
{\LARGE \bf Results from the analysis of data collected with 50 m$^2$ RPC 
carpet at the YangBaJing Laboratory}
\end{center}

\begin{center}
%
%
{\bf ARGO-YBJ Collaboration}, presented by B. D'Ettorre Piazzoli$^{1}$\\
{\it $^{1}$ INFN and Dipartimento di Scienze Fisiche dell'Universit\'a, Napoli 
(Italy)}
\end{center}

\begin{center}
{\large \bf Abstract\\}
\end{center}
\vspace{-0.5ex}
%
%
An RPC carpet covering $\sim$ 10$^4$ m$^2$ (ARGO-YBJ experiment) will be 
installed in the YangBaJing Laboratory (Tibet, China) at an altitude of 
4300 m a.s.l. . A test-module of $\sim$ 50 m$^2$ has been put in operation 
in this laboratory and about 10$^6$ air shower events have been collected. 
The carpet capability of reconstructing the shower features is presented.
\vspace{1ex}

%
%
\section{Introduction}
\label{intro.sec}

As an Italian-Chinese collaboration project, the experiment ARGO-YBJ 
(Astrophysical Radiation with Ground-based Observatory at YangBaJing) 
is under way over the next few years at Yangbajing High Altitude Cosmic Ray 
Laboratory (4300 m a.s.l., 606 $g/cm^2$), 90 km North to Lhasa 
(Tibet, P.R. China). 
The aim of the experiment is the study of cosmic rays, mainly 
cosmic $\gamma$-radiation, at an energy threshold of $\sim 100$ $GeV$, by 
means of the detection of small size air showers at high altitude. 

The apparatus consists of a full coverage detector of dimension 
$\sim 71\times 74\>m^2$ realized with a single layer of Resistive Plate 
Counters (RPCs). The area surrounding the central detector core, up to $\sim
100\times 100\>m^2$, consists in a guard ring partially ($\sim 50\> \%$) 
instrumented with RPCs. 
These outer detector improves the apparatus performance, enlarging the 
fiducial area, for the detection of showers with the core outside the 
full coverage carpet. 
A lead converter $0.5$ $cm$ thick will cover uniformly the RPC plane in order
to increase the number of charged particles by conversion of shower photons 
and to reduce the time spread of the shower front.
The site coordinates (longitude $90^{\circ}$ 31' 50'' E, latitude 
$30^{\circ}$ 06' 38'' N) permits the monitoring of the Northern hemisphere in 
the declination band $-10^{\circ}<\delta <70^{\circ}$. 

ARGO-YBJ will image with high sensitivity atmospheric showers induced by
photons with $E_{\gamma}\geq 100$ $GeV$, allowing to bridge the
GeV and TeV energy regions and to face a wide range of fundamental issues in
Cosmic Ray and Astroparticle Physics (Abbrescia et al. (1996)):
\begin{verse}
1) {\it \ul {Gamma--Ray Astronomy}}, at a $\sim 100$ $GeV$ threshold energy. 
Several galactic and extragalactic point candidate sources can be 
continuously monitored, with a sensitivity to unidentified sources better than 
$10\%$ of the Crab flux.\\
2) {\it \ul {Diffuse Gamma--Rays}} from the Galactic plane, 
molecular clouds and SuperNova Remnants at energies $\geq 100\>GeV$.\\
3) {\it \ul {Gamma Ray Burst physics}}, by allowing the extension of the 
satellite measurements over the full GeV/TeV energy range.\\
4) {\it \ul {$\overline{p}/p$ ratio}} at energies $300\>GeV\div TeV$ not 
accessible to satellites, with a sensitivity adequate to distinguish 
between models of galactic or extragalactic $\overline{p}$ origin.\\
5) {\it \ul {Sun and Heliosphere physics}}, including cosmic ray 
modulations at  $10\>GeV$ threshold energy, the continuous 
monitoring of the large scale structure of the interplanetary 
magnetic field and high energy gamma and neutron flares from the Sun.
\end{verse}

Additional items come from using ARGO-YBJ as a traditional EAS array 
covering the full energy range from $10^{11}$ to $10^{16}\>eV$. 
Since the detector provides a high granularity space-time picture of the 
shower front, detailed study of shower properties as, for instance, 
multicore events, time and lateral distributions of EAS particles, 
multifractal structure of particle densities near the core, can be 
performed with unprecedented resolution. 
Detector assembling will start late in 2000 and data taking with the first 
$\sim$ 750 $m^2$ of RPCs in 2001. 

In order to investigate both the RPCs performance at 4300 $m$ a.s.l. and 
the capability of the detector to sample the shower front of atmospheric 
cascades, during 1998 for the first time a full coverage carpet of 
$\sim 50\>m^2$ has been put in operation in the Yangbajing Laboratory. 
The present paper is a report of this test-experiment. 

\section{The test experiment at Yangbajing}
\label{testexp.sec}

The basic elements of ARGO-YBJ detector are RPCs of dimensions $280\times 
125\>cm^2$. The detector is organized in modules of 12 chambers whose 
dimensions are $5.7\times 7.9\>m^2$. 
This group of RPCs  represent a logical subdivision (Cluster) of the 
apparatus: the detector consists of 117 Clusters in the central part and 
28 Cluster in the guard ring for a total of 1740 RPCs. The proposed 
lay-out allows to achieve an active area of $\sim 92\%$ the total. 
The trigger and the DAQ systems are built following a two level
architecture. The signals from the Cluster are managed by Local Stations.
The information from each Local Station is collected and elaborated in the
Central Station. According to this logic a module of 12 RPCs (i.e. the Cluster)
represents the basic detection unit.

A Cluster prototype, very similar to that proposed for ARGO-YBJ 
experiment, has been installed in the Yangbajing Laboratory. It 
consists of 15 chambers distributed in 5 columns with an active area of 
$\sim 90\%$. The total area is about $6.1\times 8.7$ $m^2$. 

The detector consists of single-gap RPCs made of bakelite (with volume 
resistivity 
$\rho > 5\cdot 10^{11}\>\Omega\cdot cm$) with a $280\times 112\>cm^2$ area. 
The RPCs read-out is performed by means of Al strips 3.3 $cm$ wide and 56 
$cm$ long at the edge of which the front-end electronics is connected.
The FAST-OR of 16 strips defines a logical unit called pad: ten pads 
($56\times 56\>cm^2$) cover each chamber. The FAST-OR signal from each pad 
is sent, via coaxial cable, to dedicated modules that generate the trigger 
and the STOP signals to the TDCs. Each channel of the TDCs 
measures, with $1$ $ns$ clock, the arrival times (up to 
16 hits per channel) of the particles hitting a pad, for a total number 
of 150 timing channels. 
The RPCs have been operated in streamer mode with a gas mixture of 
argon ($15\%$), isobutane ($10\%$) and tetrafluoroethane $C_2H_2F_4$ 
($75\%$), at a voltage of 7400 $V$, about 500 $V$ above the plateau knee. 
The efficiency of the detector, as measured by a small telescope selecting 
a cosmic ray beam, is $>95\%$, and the intrinsic time resolution 
$\sigma_t\sim 1$ $ns$. The description of the results concerning the RPCs
performance at YangBaJing are given in Bacci et al. (1999). 

\section{Data Analysis}
\label{datana.sec}

Different triggers based on pad multiplicity have been used to collect 
$\sim 10^6$ shower events with $0.5$ $cm$ of lead on the whole carpet in 
April-May 1998.
The integral rate as a function of the pad multiplicity is shown in Fig. 1 
for showers before and after the lead was installed. 
A comparison at fixed rate indicates an increasing of pad multiplicity due to 
the effect of the lead of $\sim 15\div 20\%$, as expected according to our 
simulations. 

\begin{figure}[htb]
\vfill \begin{minipage}{.47\linewidth}
\begin{center}
\mbox{\epsfig{file=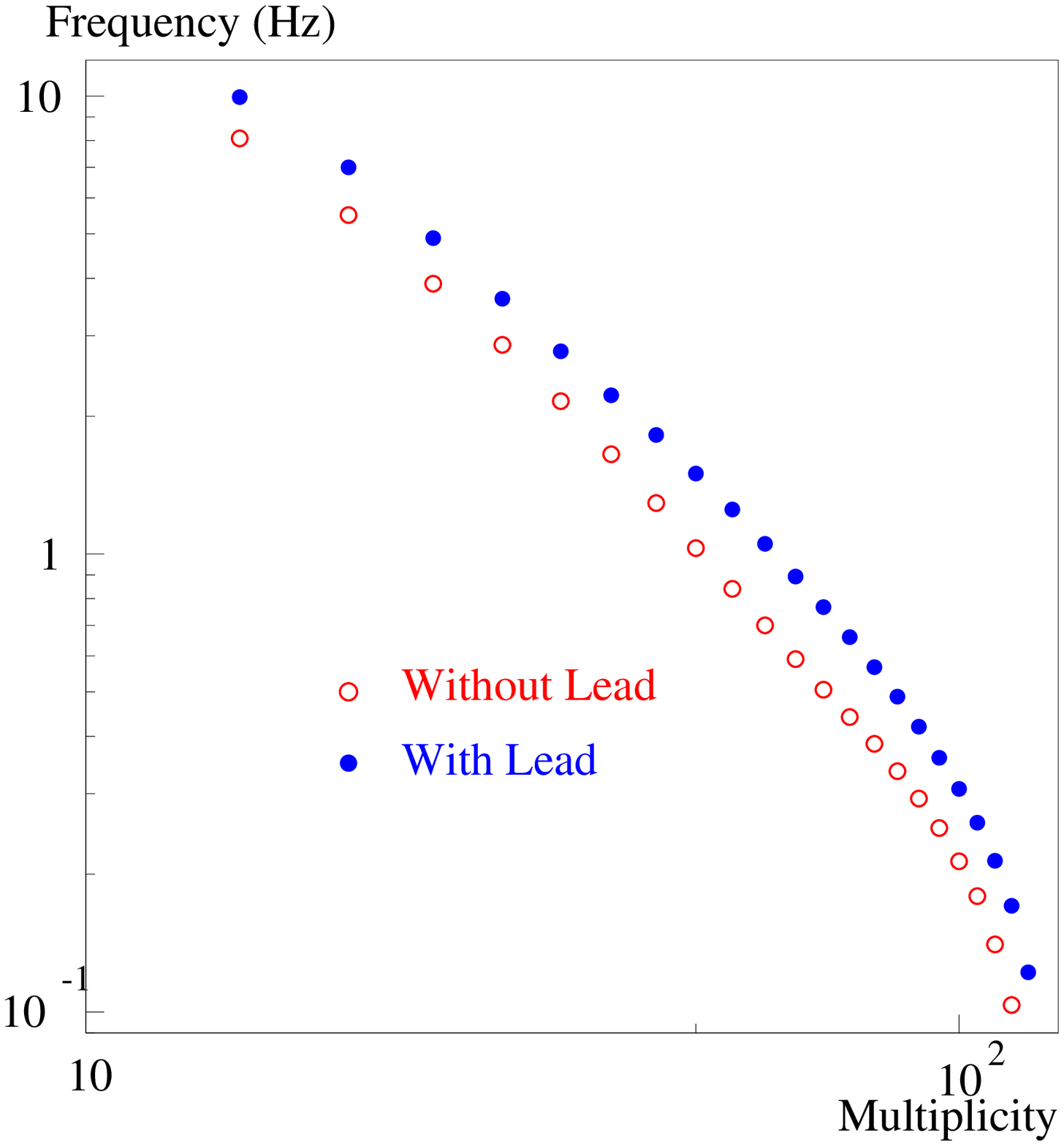,height=8.cm,width=8.cm}}
\end{center}
\caption{\em The integral rate as a function of the pad multiplicity. }
\end{minipage}\hfill
\hspace{-0.5cm}
\begin{minipage}{.47\linewidth}
\begin{center}
\mbox{\epsfig{file=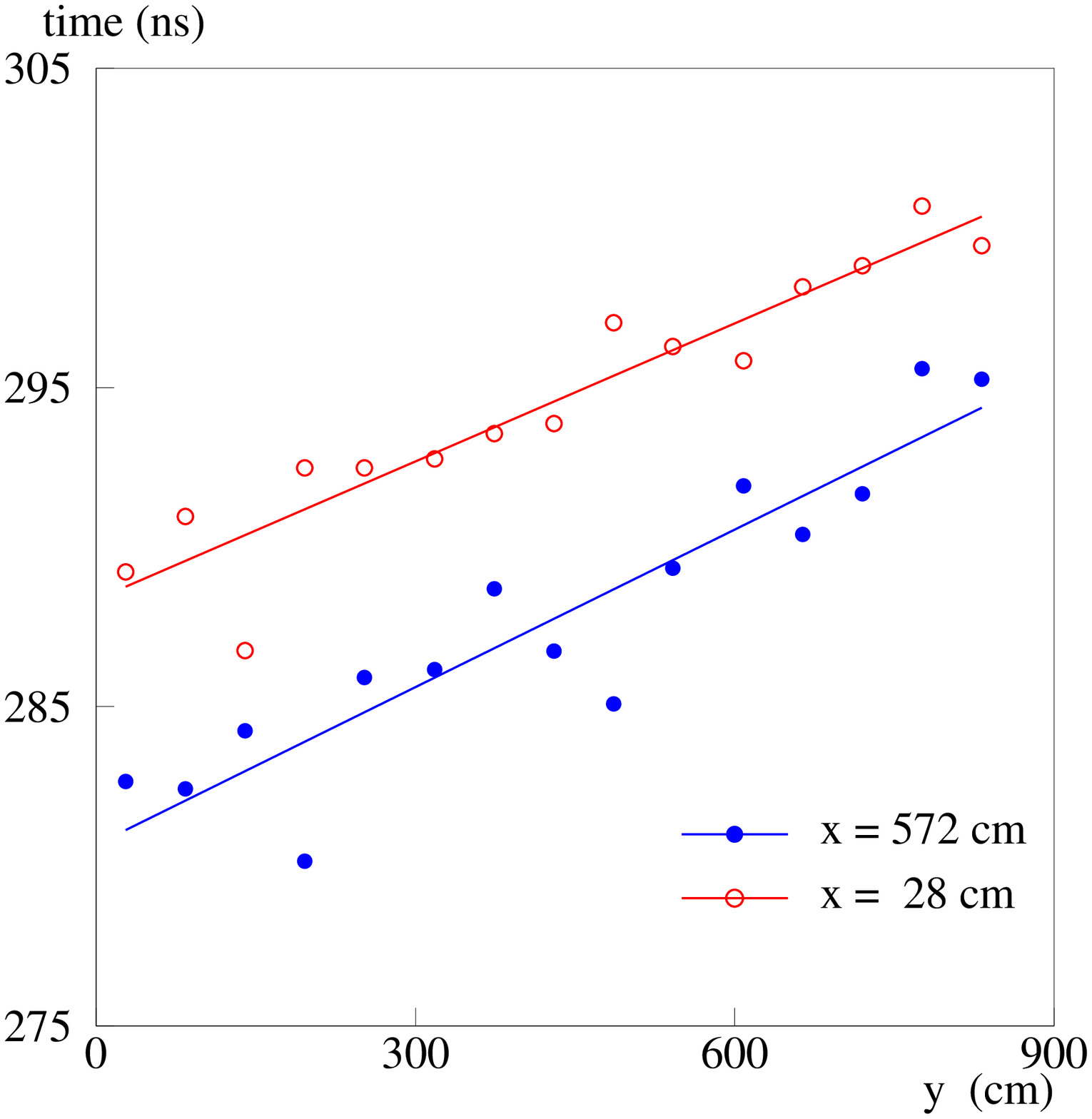,height=8.cm,width=8.cm}}
\end{center}
\caption{\em Time profile observed in a typical event. Straight lines are 
fit to experimental hits. }
\end{minipage}\hfill
\end{figure}

\subsection{Detector Calibration}
\label{detcali.sec}

The relative time offset among different pads are measured as follows: 
\begin{verse}
1) We construct the time distributions of each TDC channel adding all 
the delays in the individuals events and compute the relative time 
mean values $\ol{t_i}$. \\
2) We compute the mean value of the $\ol{t_i}$ distribution: 
$<t>={ { \sum_{i=1}^{N} \ol{t_i} } \over N}$.\\
3) We define the time offset as $\Delta t_i=\ol{t_i}-<t>$. These values 
are used to correct the times provided by each TDC channel. 
\end{verse}
To check the consistency of the procedure, we fit on an event-by-event 
basis the corrected time values to a plane, construct for each TDC channel 
the distribution of time residuals $\delta t_i=t_{plane}-t_i$ and calculate 
the mean values $<\delta t_i>$. These are distributed with a spread of 
$\sim$ $0.6$ $ns$. 

\subsection{ Event Reconstruction}

In this test-experiment we are not able to determine the core position, 
therefore we use a plane as the fitting function. Moreover, the estimated 
arrival direction is relatively free from the curvature effect because we 
sample only a small portion of the shower front. 
In this approximation the expected particle arrival time is a linear 
function of the position.
The time profile observed in a typical event is shown in Fig. 2. Here $x,y$ 
are orthogonal coordinates which identify the pad position. Straight 
lines are one-dimensional fits to experimental hits along two different $x$ 
values. 

We performe an optimized reconstruction procedure of the shower direction 
as follows:
\begin{verse}
1) Unweighted plane fit to hits for each event with pad multiplicity 
$\geq$ 25. \\ 
2) Rejection of out-lying points by means of a 2.5 $\sigma$ cut and iteration 
of the fit until this condition is not verified.\\
\end{verse}
After these iterations a fraction $\leq 10\%$ of the time signals that
deviate most from the fitted plane are excluded from further analysis. 
The distribution of time residuals $\delta t$ = $t_{plane}-t_i$
(Fig. 3) exhibits a long tail due to time fluctuations and to the
curved profile of the shower front, more pronounced for low multiplicity. 
The width of these distributions is related to the time thickness of the 
shower front. Since the position of the shower core is not reconstructed, 
the experimental result concerns a time thickness averaged on different radial 
distances. Increasing pad multiplicity select showers with core near the 
detector, as confirmed by MC simulations. Taking into account the total 
detector jitter of $1.3$ $ns$ (RPC intrinsic jitter, strip length, 
electronics time resolution) the time jitter of the earliest particles in 
high multiplicity events ($116\div 120$ hits) is estimated $\sim 1.1$ $ns$.

\begin{figure}[htb]
\vfill \begin{minipage}{.47\linewidth}
\begin{center}
\mbox{\epsfig{file=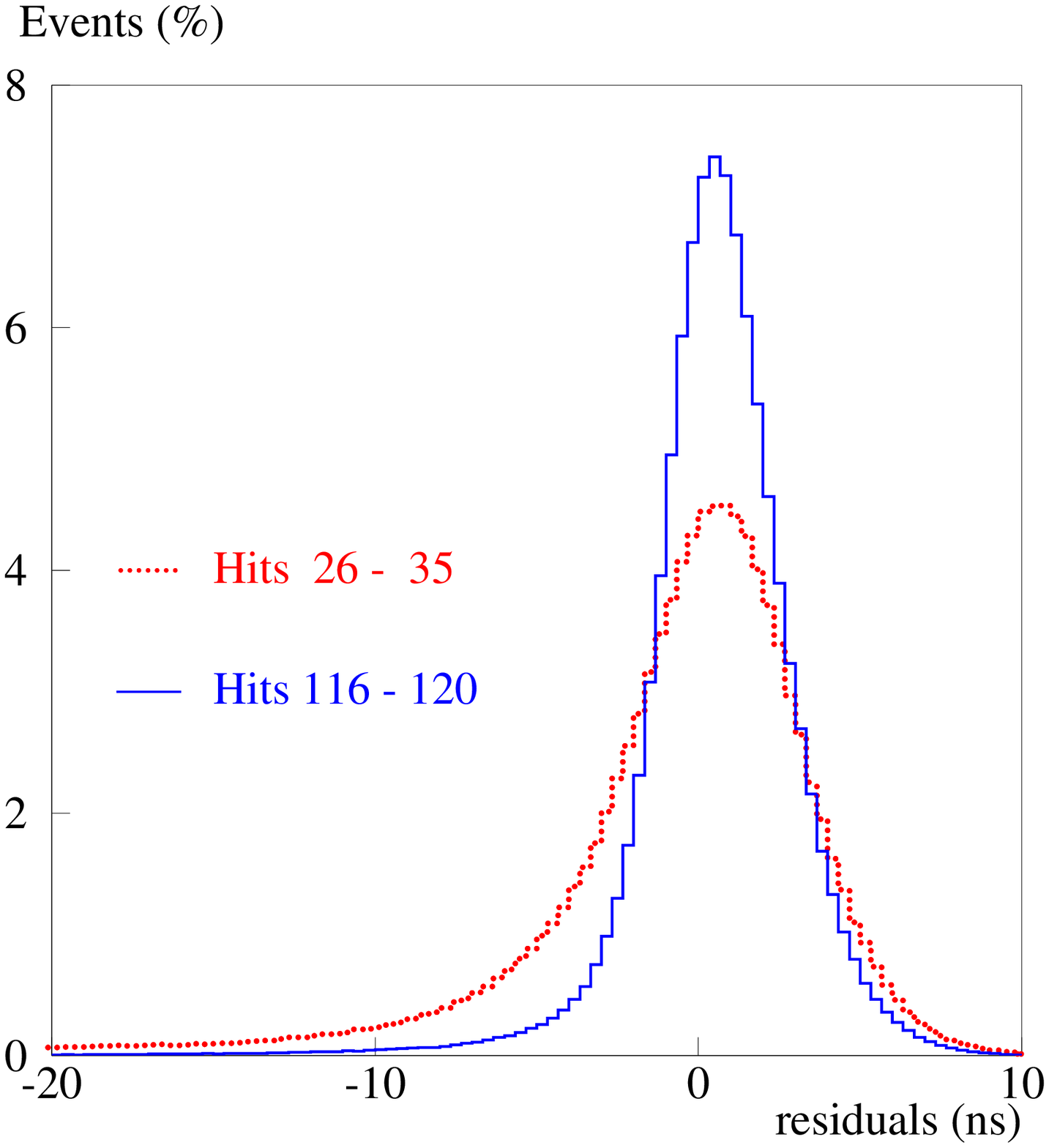,height=8.cm,width=8.cm}}
\end{center}
\caption{\em Distribution of time residuals for events with different pad 
multiplicity (all channel added).}
\end{minipage}\hfill
\hspace{-0.5cm}
\begin{minipage}{.47\linewidth}
\begin{center}
\mbox{\epsfig{file=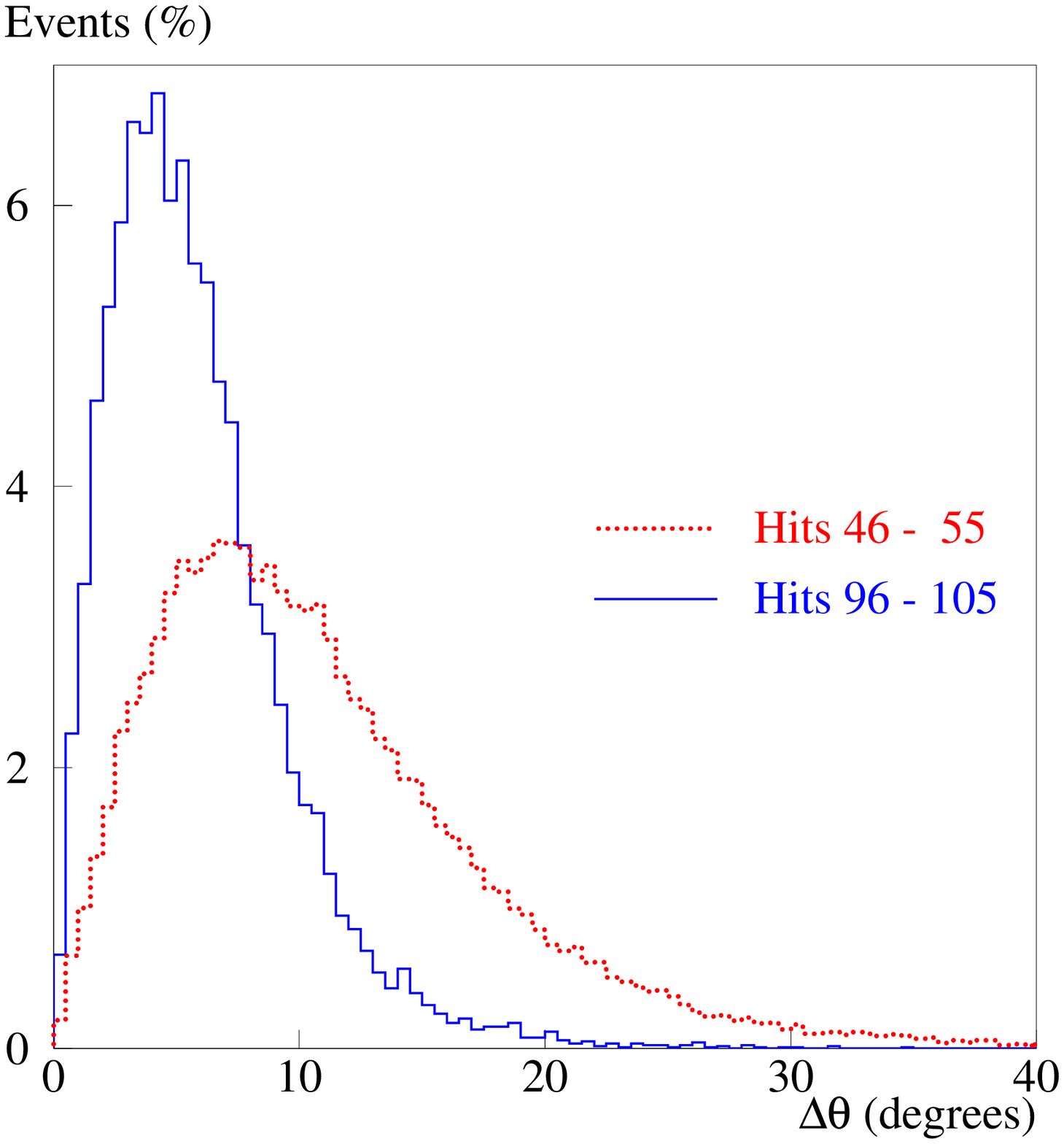,height=8.cm,width=8.cm}}
\end{center}
\caption{\em Even-odd angle difference distribution for events with 
different pad multiplicity.}
\end{minipage}\hfill
\end{figure}

\subsection{Angular Resolution}

The angular resolution of the carpet has been estimated by dividing 
the detector into two independent sub-arrays and comparing the two
reconstructed shower directions. 
Events with N total hits have been selected according to the 
constraint $N_{odd}\simeq N_{even}\simeq N/2$. 
The even-odd angle difference $\Delta \theta_{eo}$ is shown in Fig. 4 for 
events in different multiplicity ranges. 
We note that these distributions narrow, as expected, with the increase of 
the shower size. 
To see the dependence of the angular resolution on the lead sheet, we show 
in Fig. 5 the median $M_{\Delta \theta_{eo}}$ of the distribution of 
$\Delta \theta_{eo}$ as a function of pad multiplicity for showers 
reconstructed before and after the lead was added. 
The improvement of the angular resolution is a factor $\sim$ 1.4 for $N=50$
and decreases with increasing multiplicity. 

Assuming that the angular resolution for the entire array is Gaussian 
(Alexandreas et al. (1992)), a standard deviation 
$\sigma_{\theta}\sim 2.1^{\circ}$ is found for events with a pad 
multiplicity $\geq 100$.

\section{Conclusions}
\label{conclu.sec}

\begin{figwindow}[1,r,%
{\mbox{\epsfig{file=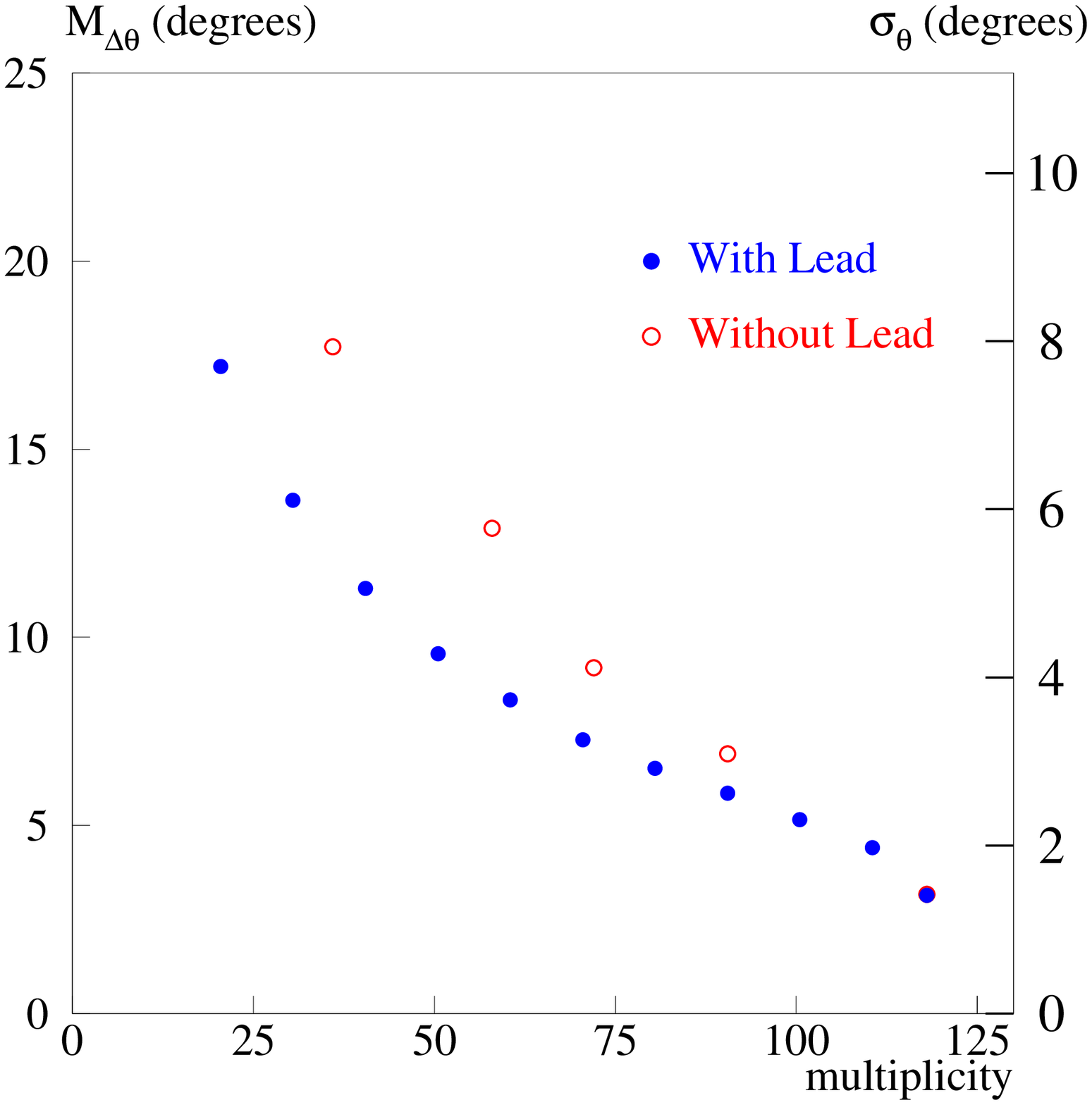,height=7.cm,width=7.cm}}},%
{\em Median of $\Delta \theta_{eo}$ distribution as a function of
pad multiplicity.}]
A Resistive Plate Counters carpet of $\sim$ 50 $m^2$ has been put in 
operation at the Yangbajing Laboratory to study the high altitude 
performance of RPCs and the detector capability of imaging with high 
granularity a small portion of the EAS disc, in view of an enlarged use 
in Tibet (ARGO-YBJ experiment). 

In this paper we have presented the results of this test experiment 
concerning the carpet capability of reconstructing the shower features. 
In particular, we have focused on the angular resolution 
in determining the arrival direction of air showers, the most important 
parameter for $\gamma$-ray astronomy studies. 

The effect of a $0.5$ $cm$ lead sheet on the whole carpet has been 
investigated. An increase $15\div 20\%$ of the hit multiplicity is found. 
The improvement of the angular resolution depends on the shower density. 

The test confirms that RPCs can be operated efficiently to sample air showers 
at high altitude with excellent space and time resolutions. 
The results are consistent with data assumed in the computation of the 
performance of the ARGO-YBJ detector. 
\end{figwindow}

%
%
%
%
%
\vspace{1ex}
\begin{center}
{\Large\bf References}
\end{center}
%
Abbrescia M. et al., {\it Astroparticle Physics with ARGO}, Proposal (1996). 
This document can be downloaded at the URL: 
http://www1.na.infn.it/wsubnucl/cosm/argo/argo.html\\
Alexandreas D.E. et al., Nucl. Instr. Meth. A311 (1992) 350.\\
Bacci C. et al. (ARGO-YBJ coll.), (1999) submitted to Nucl. Instr. Meth.\\
\end{document}